\documentclass{emulateapj_helfand}
\usepackage{apjfonts}

\newcommand{\kms}   {km\,s$^{-1}$}
\newcommand{\mpy}   {mas\,yr$^{-1}$} 
\newcommand{\pmra}  {\mu_{\alpha}}
\newcommand{\pmdec} {\mu_{\delta}}
\newcommand\psr{XTE~J1810--197}

\journalinfo{The Astrophysical Journal, in press}
\slugcomment{Received 2007 January 23; accepted 2007 March 12}

\shorttitle{THE FIRST MAGNETAR PROPER MOTION}
\shortauthors{HELFAND ET AL.}

\begin{document}

\title{VLBA measurement of the transverse velocity of the magnetar
XTE~J1810--197}

\author{
  D.~J.~Helfand,\altaffilmark{1}
  S.~Chatterjee,\altaffilmark{2}
  W.~F.~Brisken,\altaffilmark{3}
  F.~Camilo,\altaffilmark{1}
  J.~Reynolds,\altaffilmark{4}
  M.~H.~van Kerkwijk,\altaffilmark{5}
  J.~P.~Halpern,\altaffilmark{1}
  and S.~M.~Ransom\altaffilmark{6}
}

\altaffiltext{1}{Columbia Astrophysics Laboratory, Columbia University,
550 West 120th Street, New York, NY~10027; djh@astro.columbia.edu,
fernando@astro.columbia.edu, jules@astro.columbia.edu}

\altaffiltext{2}{School of Physics A28, The University of Sydney,
NSW~2006, Australia; S.Chatterjee@physics.usyd.edu.au}

\altaffiltext{3}{National Radio Astronomy Observatory, PO Box 0, Socorro,
NM~87801; wbrisken@aoc.nrao.edu}

\altaffiltext{4}{Australia Telescope National Facility, CSIRO,
Parkes Observatory, P.O. Box 276, Parkes, NSW~2870, Australia;
John.Reynolds@atnf.csiro.au}

\altaffiltext{5}{Department of Astronomy and Astrophysics, University
of Toronto, 50 Saint George Street, Toronto, ON M5S~3H4, Canada;
mhvk@astro.utoronto.ca}

\altaffiltext{6}{National Radio Astronomy Observatory, 520 Edgemont Road,
Charlottesville, VA~22903; sransom@nrao.edu}

\begin{abstract}

We have obtained observations of the magnetar \psr\ with the Very Long
Baseline Array at two epochs separated by 106 days, at wavelengths of
6\,cm and 3.6\,cm.  Comparison of the positions yields a proper motion
value of $13.5 \pm 1.0$\,mas\,yr$^{-1}$ at an equatorial position angle
of $209\fdg4 \pm 2\fdg4$ (east of north). This value is consistent
with a lower-significance proper motion value derived from infrared
observations of the source over the past three years, also reported here.
Given its distance of $3.5 \pm 0.5$\,kpc, the implied transverse velocity
corrected to the local standard of rest is $212 \pm 35$\,km\,s$^{-1}$
($1\,\sigma$). The measured velocity is slightly below the average
for normal young neutron stars, indicating that the mechanism(s) of
magnetar birth need not lead to high neutron star velocities.  We also
use Australia Telescope Compact Array, Very Large Array, and these VLBA
observations to set limits on any diffuse emission associated with the
source on a variety of spatial scales, concluding that the radio emission
from \psr\ is $>96\%$ pulsed.

\end{abstract}


\section{Introduction}

Neutron stars manifest the most extreme values of many stellar parameters
including radius, density, surface temperature, rotation rate, magnetic
field strength, and velocity. The relationships among these parameters are
thought to provide constraints on such fundamental issues as the equation
of state for matter at supra-nuclear densities and the conditions at the
center of core-collapse supernovae.  The connection between the final
two properties, the inferred surface magnetic dipole field strength and
the neutron star space velocity, has been debated for more than thirty
years. Helfand \& Tademaru (1977) first suggested a bimodal distribution
of radio pulsar velocities and claimed a correlation between velocity and
magnetic field strength. As the number of measured pulsar proper motions
increased, data presented by Anderson \& Lyne (1983) and Cordes (1986)
debunked the bimodal distribution, but appeared to confirm the correlation
between velocity and magnetic field strength, and Bailes (1989) explained
it in the context of the formation of neutron stars in binary systems.
As yet larger samples became available, the bimodality of the velocity
distribution reappeared (cf. Narayan \& Ostriker 1990; Lorimer et al.\
1995; Cordes \& Chernoff 1998; Brisken et al.\ 2003) --- and once again
disappeared (Hobbs et al.\ 2005; Faucher-Gigu{\`e}re \& Kaspi 2006) ---
while the relationship between velocity and magnetic field strength was
explained as an evolutionary difference in subpopulations rather than
a physical connection between the origin of spin period, magnetic field
strength, and velocity kick at birth (Cordes \& Chernoff 1998).

Magnetars, first discussed by Duncan \& Thompson (1992), were postulated
to have the most extreme magnetic properties of all neutron stars
with mean dipole field strengths ranging from  $\sim\!10^{14}$ to
$10^{15}$\,G, more than 100 times that of ordinary pulsars.  Duncan and
Thompson suggested that these high fields would form as a consequence
of dynamo action in very rapidly rotating (spin period $\sim\!1$\,ms)
proto-neutron stars.  Magnetic braking in such stars would be extremely
rapid, leading to the prediction of young, slowly rotating stars with very
large magnetic dipole moments. The recognition of Anomalous X-ray Pulsars
(AXPs) as a distinct class of neutron stars with periods of 5--12\,s
(Mereghetti \& Stella 1995), and the confirmation of long rotation periods
as a characteristic of Soft Gamma-ray Repeaters (SGRs; Kouveliotou 1999)
led to a compelling argument, first suggested by Paczinski (1992),
that these objects had very large magnetic fields, the decay of which
powered their quiescent X-ray emission (Thompson \& Duncan 1996).

Motivated by the large offset of the SGR 0526--66 from the center of
its putative natal supernova remnant (SNR) N49 (Rothschild et al.\
1994), as well as by the since discredited notion that normal gamma-ray
bursts might originate in halo neutron stars, Duncan \& Thompson
(1992) suggested several mechanisms by which the high magnetic fields
characteristic of magnetars would also lead to very high space velocities
of $\sim\!10^3$\,km\,s$^{-1}$.  These processes include asymmetric mass
loss either at the time of collapse or in an anisotropic magnetized wind
or jet, as well as anisotropic neutrino emission and/or the photon rocket
effect.  They noted that most of these mechanisms are ineffective for the
normal pulsar population (although a few outliers of that population have
velocities in excess of $10^3$\,km\,s$^{-1}$; Chatterjee et al.\ 2005).
However, the recognition that AXPs and SGRs were both manifestations
of magnetars, and that the former were, in several cases, centrally
located in their natal SNRs, led Gaensler et al.\ (2001) to express
skepticism regarding most proposed SGR/SNR associations and to conclude
that magnetars as a class had velocities $<500$\,km\,s$^{-1}$.

Measurements of magnetar velocities thus provide an important clue
regarding the AXP/SGR connection, as well as offering a diagnostic of
the events accompanying the birth of magnetars including the generation
of their extraordinary magnetic fields.  Proper motion measurements are
thus highly desirable.  The high level of timing noise present in all
magnetars with phase-connected timing solutions (see Woods \& Thompson
2006, for a review, and Camilo et al.\ 2007a) renders a timing proper
motion measurement impossible.  To date, direct measurements have been
prevented by the fact that the handful of known magnetars are distant,
and are primarily detected at X-ray wavelengths where the angular
resolution of the best telescope, {\em Chandra}, is $0\farcs5$; the few
optical/infrared (IR) detections are very faint (e.g., Hulleman et al.\
2000), making direct proper motion measurements difficult.  Nonetheless,
attempts are underway to measure proper motions in both of these
wavelength regimes; in \S~3 we describe such a measurement in IR for \psr.

The transient AXP \psr, with a rotation period of 5.54\,s, was discovered
following an increase in X-ray brightness by a factor of $>100$ in early
2003 (Ibrahim et al.\ 2004), after which it began an exponential decay,
heading to its pre-outburst level by 2007 (Gotthelf \& Halpern 2007).
It was then detected at IR wavelengths (Israel et al.\ 2004; Rea et
al.\ 2004).  Subsequently, Halpern et al.\ (2005) discovered that \psr\
was coincident with a radio source in the MAGPIS survey of the Galactic
plane at 20\,cm (Helfand et al.\ 2006).  Followup observations with the
Parkes telescope revealed \psr\ to be a bright radio pulsar with a number
of unique properties (Camilo et al.\ 2006) including a flat spectrum,
highly variable flux densities and pulse shapes at all frequencies
from 0.3\,GHz to 144\,GHz and on all timescales from minutes to months,
as well as having a very high degree of linear polarization ($>95\%$;
Camilo et al.\ 2007b).

The radio emission from \psr\ opens the possibility of using the Very
Long Baseline Array (VLBA) to obtain the first magnetar proper motion. We
report here our two-epoch measurement of the position of \psr\ which
leads to the first measurement of a magnetar transverse velocity. In
Section 2 we describe the VLBA observations, their reduction, and the
proper motion determination.  Section 3 describes a consistent IR proper
motion measurement.  Section 4 uses data from the Australia Telescope
Compact Array (ATCA) and the Very Large Array (VLA), as well as the
VLBA observations to set limits on any synchrotron nebula surrounding
the magnetar. The final section summarizes the distance determinations
for \psr, derives the source velocity, and comments on the implications
of the low value we find for the origin of magnetars.

\section{VLBA Observations and Proper Motion of XTE~J1810--197}

We observed \psr\ with the NRAO VLBA on 2006 June 2 under an Exploratory
Time request.  While most radio pulsars are observed at 1.4\,GHz,
we took advantage of the extraordinarily flat spectrum of \psr\
($S_\nu \propto \nu^{-\alpha}, \; \alpha \sim\!0$; Camilo et al.\
2006) to observe at 5\,GHz and 8.4\,GHz, where the VLBA offers both
higher spatial resolution and better system temperatures; in addition,
ionosphere-induced astrometric errors scale as $\nu^{-2}$, and are
thus greatly reduced at these higher frequencies, a benefit especially
important for this low-declination source which must always be observed
at low elevation angles.  As is usual in astrometric projects where
absolute positional information is required, the observations were
phase-referenced by nodding back and forth between the target and a
reference calibrator source, J1753--1843, which lies $\sim\!4\arcdeg$
away.  We used a cycle time of 90\,s on the calibrator and 120\,s on
the target to ensure that visibility phase coherence was preserved.
The bandwidth at both frequencies was 32\,MHz for each of the L and R
circular polarizations. The L and R correlation products were summed
at imaging time to form total intensity.  Linear polarization products
(i.e., L $\times$ R) were not generated.  A total of 58 minutes of
on-source data was accumulated in each band using 2\,s integration times.

While proper motion measurements benefit from long time baselines,
the rapid decline in flux density observed for \psr\ (Camilo et al.\
2007a) led us to acquire a second-epoch observation only 106 days later,
on 2006 September 16, using identical observational parameters.

At both epochs and both frequencies, the VLBA correlator was gated using a
pulse timing solution determined from observations at Nan\c{c}ay, Parkes,
and the GBT (Camilo et al.\ 2007a) in order to boost the signal-to-noise
ratio (S/N) for the magnetar.  We employed a boxcar gate with a width of
4\% of the pulse period, resulting in a S/N boost of $0.04^{-1/2} = 5$
(assuming all of the radio pulsed flux is captured in the gate window).
Given the timing noise and pulse shape variability, especially at later
epochs (Camilo et al.\ 2007a), the fraction of the pulsed emission
actually captured in the gate is somewhat uncertain, although for the
first epoch, the off-pulse image shows no sign of the source (see \S~4.3).

The correlated data were reduced in AIPS using standard procedures.
The visibility amplitudes were calibrated based on the measured system
temperature at each antenna, and the target visibility phases, rates,
and delays, as well as the bandpass responses, were calibrated using
the observed visibility phases of the reference calibrator.  We also
included corrections for ionospheric propagation effects based on
global models of the total electron content (TEC) in the Earth's
ionosphere, which are derived every 2 hours from dual-frequency
Global Positioning System measurements around the world\footnote{See
ftp://cddis.gsfc.nasa.gov/gps/products/ionex .}.  Finally, the calibrated
visibilities were transformed to create images, and astrometric positions
were obtained at 5\,GHz and 8.4\,GHz by fitting Gaussians to the observed
point source images.  We detected \psr\ at both frequency bands and at
both epochs; the first epoch detections were strong ($\mbox{S/N}\sim
75$--150), while the second epoch detections had substantially lower
$\mbox{S/N}\sim 6$--7.

Along with the statistical uncertainty of these position measurements, two
primary sources of systematic astrometric error remain in the calibrated
data.  First, the position of the reference calibrator J1753--1843 is
known only to a precision of 4\,mas (Petrov et al.\ 2006), which results
not only in an absolute position offset (inconsequential for a proper
motion measurement), but also increases slightly ($\sim\!0.1$\,mas)
the systematic errors in the data.  Second, the southern declination of
\psr\ implies a low elevation at all VLBA antennas, leading to increased
atmospheric and ionospheric path lengths.  Combined with the 4\arcdeg\
separation between the calibrator and the target, this produces a
substantial uncalibrated path differential, which is only partially
corrected by the global TEC models.  As illustrated in Chatterjee et al.\
(2004, Fig. 3), we expect the path differential to contribute systematic
errors of $\sim\!1$\,mas to the astrometry at 5\,GHz.  Note that the
angular separation between the target and calibrator is $4\arcdeg$ in
right ascension and $1\arcdeg$ in declination, and the mismatch between
positions derived at 5\,GHz and 8.4\,GHz is also substantially worse in
right ascension, consistent with an effect similar to that observed by
Chatterjee et al.\ (2004).  Thus, even though the image S/N is highest
for the first-epoch 5\,GHz observations, the first-epoch 8.4\,GHz data
is less affected by residual ionospheric effects and leads to a better
absolute position for \psr, as listed in Table~\ref{Table:fit}.

\begin{deluxetable}{ll}
\tablewidth{0.80\linewidth} 
\tablecaption{\label{Table:fit} Parameters for \psr\ }
\tablecolumns{2}
\tablehead{ 
\colhead{Parameter} & \colhead{Value}
}
\startdata 
Epoch\dotfill                   & 2006.42 (MJD~53888)                         \\
Right Ascension, $\alpha_0$\dotfill & $18^{\mathrm h} 09^{\mathrm m} 51\fs08696$ \\
Declination, $\delta_0$\dotfill & $-19\arcdeg 43\arcmin 51\farcs9315$         \\
$\pmra$ (\mpy)\dotfill          & $-6.60 \pm 0.06$                            \\
$\pmdec$ (\mpy)\dotfill         & $-11.72 \pm 1.03$                           \\
                                &                                             \\
Proper motion, $\mu$ (\mpy)\dotfill & $13.5 \pm 1.0$                          \\
Equatorial p.a. (east of north) & $209\fdg4 \pm 2\fdg4$                       \\
Distance (kpc)\dotfill          & $3.5\pm0.5$                                 \\
$V_{\perp,{\rm LSR}}$\tablenotemark{a} (\kms)\dotfill & $212\pm35$            \\
Galactic coordinates, $l,b$\dotfill &  10\fdg73, --0\fdg16
\enddata 
\tablecomments{All astrometric parameters are in the J2000 equinox.
Absolute positions are referenced to J1753--1843 ($\mbox{R.A.} =
17^{\mathrm h} 53^{\mathrm m} 09\fs088754$, $\mbox{Decl.} = -18\arcdeg
43' 38\farcs52316$), which has a positional uncertainty of 4\,mas.
Our reported position for \psr\ is from the 8.4\,GHz data at the first
epoch, which has the lowest systematic errors, but the uncertainties
encompass the position measured at 5\,GHz as well (see \S~2). Errors
quoted are $1\,\sigma$.
}
\tablenotetext{a}{This velocity in the plane of the sky has been corrected
to the local standard of rest using the distance given, and values for
the Galactic rotation velocity (assumed constant) and peculiar velocity
of the Sun from Cox (2000). }
\end{deluxetable} 

\begin{figure}[h]
\plottwo{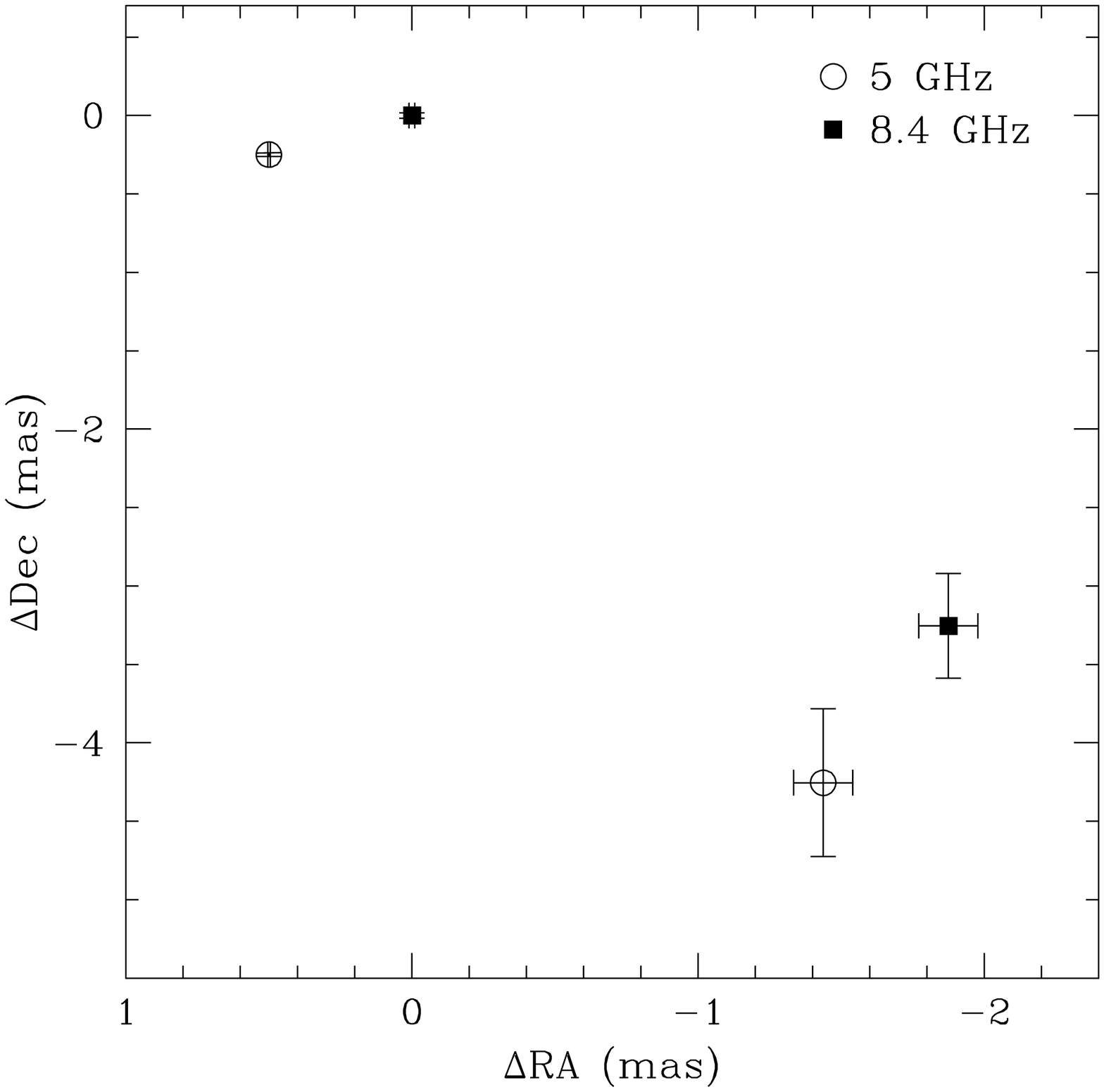}{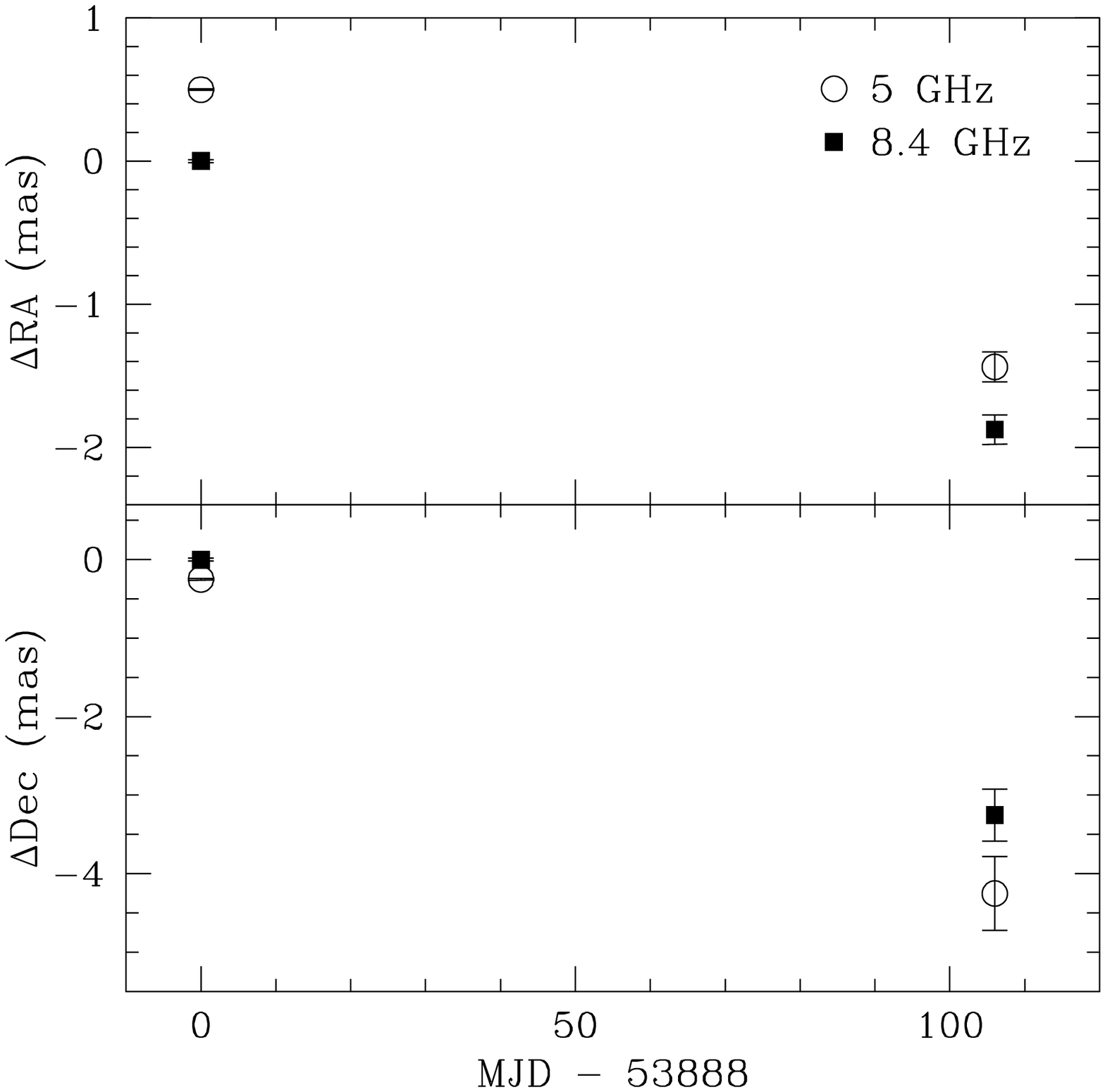}
\caption{\label{Fig:pm}  The proper motion of \psr\ using VLBA
  measurements. \textit{Left}: Position offsets in right ascension
  and declination on MJD~53888 and MJD~53994, as measured with the VLBA at
  5\,GHz (\textit{circles}) and 8.4\,GHz (\textit{solid squares}). Offsets
  are measured with respect to the 8.4\,GHz position on MJD\,53888,
  as listed in Table~\ref{Table:fit}.  \textit{Right}: The same
  position offsets, shown as a function of time.  The plotted error
  bars represent only the random position uncertainties, while the
  offset between the positions at different observing bands on the same
  day illustrates the residual systematic errors in the astrometry.
  The decrease in source flux density over time results in an increase
  in the random position error, and hence larger error bars, at the
  second epoch.
}
\end{figure}

For a proper motion measurement, only the relative displacement (as
shown in Fig.~\ref{Fig:pm}) is of importance, and so we fit independent
proper motions to the positions in the two frequency bands.  While this
precludes a goodness of fit estimate, our measurements are dominated
by systematic errors in any case. At 8.4\,GHz, we measure a proper
motion $\pmra = -6.54$\,\mpy\ and $\pmdec = -11.20$\,\mpy, while at
5\,GHz, we measure $\pmra = -6.66$\,\mpy\ and $\pmdec = -13.78$\,\mpy.
We combine these measurements as a weighted average, estimate our residual
systematic errors from the scatter between the measurements, and add them
in quadrature with the (smaller) statistical errors to derive our final
uncertainties in the proper motion, as reported in Table~\ref{Table:fit}.

\section{Motion of the Infrared Counterpart}

The IR source identified by Israel et al.\ (2004) on the basis of precise
coincidence with the {\em Chandra\/} position and unusual colors is
demonstrably the counterpart of \psr\ because of its variability (Rea et
al.\ 2004).  To the extent that we can measure it, the proper motion of
the IR counterpart should therefore agree with the VLBA result.  Here,
we show that adaptive optics observations allow a proper motion to be
measured, and that it is consistent.  This method is therefore feasible
for measuring or significantly constraining the proper motions of other
magnetars, all of which are (thus far) radio quiet.

For this purpose, we used new and archival IR $K$-band observations.
Specifically, we used a relatively wide-field image obtained with
the near-IR imager NIRI on Gemini in 2003 September ($2\times2\,{\rm
arcmin}^2$), an image obtained with the adaptive optics system Altair
on Gemini in 2006 September ($22\times22\,{\rm arcsec}^2$), and two
images obtained with the adaptive optics system NACO on the Very Large
Telescope in 2004 March and September ($27\times27\,{\rm arcsec}^2$).
The reduction of these data followed basic steps, described in more
detail in Camilo et al.\ (2007, in preparation).

We tied our observations to the International Celestial Reference System
(ICRS) using 66 stars from the 2MASS catalog (Skrutskie et al.\ 2006) that
had single, stellar counterparts on the NIRI image.  Rejecting 8 outliers,
the root-mean-square (rms) residuals per star were $\sim\!80$\,mas in each
coordinate, consistent with the 2MASS measurement uncertainties.  Using 37
fainter stars, we transferred this solution to the Altair image (with
residuals per star of $\sim\!23\,$mas).  Given the residuals in the two
steps, our observations are tied to the 2MASS catalog to $\sim\!10\,$mas.
We found that the counterpart was offset from the VLBA position derived
in \S~2 by $\Delta\alpha=-0\farcs00$ and $\Delta\delta=-0\farcs05$,
which is well within the systematic uncertainty of $\la\!0\farcs1$
with which the 2MASS catalog is tied to the ICRS (Skrutskie et al.\ 2006).

\begin{figure}
\epsscale{0.7}
\plotone{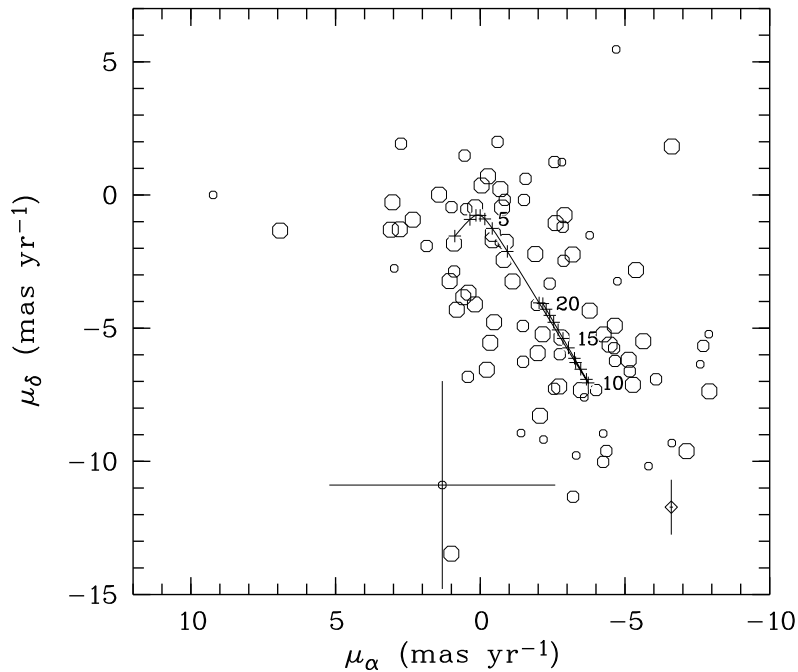}
\caption{\label{fig:irpm} The proper motion of \psr\ (circle with
  error bars) derived from infrared observations relative to those
  for stars in the field (see \S~3).  The symbol sizes correspond
  to magnitude ranges, with, from large to small, $K<20$, $20<K<21$,
  and $K>21$, corresponding to proper-motion uncertainties of 0.7--1,
  1--2, and 2.1--4${\rm\,mas\,yr^{-1}}$, respectively.  The proper
  motions were placed on an absolute scale by shifting them such that
  they matched the proper motions expected due to Galactic rotation
  (with an assumed constant velocity of $220{\rm\,km\,s^{-1}}$) and solar
  peculiar velocity.  These expectations are shown by the drawn curve for
  distances from 1\,kpc to 20\,kpc. The VLBA proper motion measurement
  (\S~2; Fig.~\ref{Fig:pm}) is shown by the diamond with error bars
  toward the bottom right of the plot.
}
\end{figure}

To determine proper motions, we measured centroids for 93 stars
common to the three adaptive optics images.  The proper motions thus
measured depended in a systematic way on position, due to distortion.
We corrected for differences in distortion in the different cameras
by first obtaining quadratic, six-parameter fits for the proper
motions of the stars as a function of offset in right ascension and
declination relative to the position of \psr.  We then subtracted
the fit corrections from the measured proper motions (this procedure
ensures that the proper motion of \psr\ at this stage is relative to
those of stars near it on the detectors).  In order to place these
proper motions on an absolute scale, we calculated the expected motions
in this direction due to Galactic rotation and the peculiar motion of
the Sun.  We find that the former dominates for distances in excess
of 1\,kpc, and leads to proper motions in Galactic longitude ranging
from $-1$ to $-8{\rm\,mas\,yr^{-1}}$ between 1\,kpc and 10\,kpc, and
declining thereafter.  Roughly matching our observed relative proper
motions to this (by applying a correction of $-4{\rm\,mas\,yr^{-1}}$
in longitude and $-0.5{\rm\,mas\,yr^{-1}}$ in latitude, and converting
back to equatorial coordinates), we infer a proper motion for the
counterpart of \psr\ of $\mu_\alpha=1\pm4{\rm\,mas\,yr^{-1}}$ and
$\mu_\delta=-11\pm4{\rm\,mas\,yr^{-1}}$ (see Fig.~\ref{fig:irpm}),
where the quoted uncertainties are measurement errors and the systematic
uncertainty is about $1{\rm\,mas\,yr^{-1}}$.  This is consistent at
the 87\% confidence level with the proper motion measured by VLBA
(see Table~\ref{Table:fit}).

\section{Limits on Extended Radio Emission}

In the paper reporting the discovery of radio emission from \psr,
Halpern et al.\ (2005) argued, based on the source's non-detection in the
Parkes multibeam pulsar survey (e.g., Manchester et al.\ 2001) and the
tight limits on radio pulsations from other magnetars, that the radio
source was likely arising as a result of diffuse synchrotron  emission
analogous to the wind nebulae seen around many young pulsars. While the
detection of radio pulses at the X-ray rotation period demonstrated that
we are presented with a new phenomenon, the presence of pulses does not
rule out the existence of diffuse emission. We have thus explored this
possibility using three separate instruments to probe for emission on
linear scales from 5\,AU to 0.5\,pc.

\subsection{ATCA Observations}

\psr\ was observed with the ATCA in its 6D configuration simultaneously
at 13\,cm (2.368\,GHz) and 20\,cm (1.384\,GHz) on 2006 June 8 under
a Target of Opportunity request.  At each frequency, for each of two
orthogonal linear polarizations, we sampled a bandwidth of 128\,MHz split
into 33 frequency channels.  The array was used in a time-binning mode
with visibilities recorded in 32 phase bins across the 5.54\,s period.
First, we observed the flux calibrator 1934--638.  We then obtained
data for approximately 7.5\,hr, with scans of $\sim\!20$\,min on the
pulsar interspersed with $\sim\!2$\,min scans on the phase calibrators
1730--130 and 1921--293.

The correlated data were analyzed using standard techniques with MIRIAD.
In order to reduce the effects of confusion from the Galaxy, we discarded
data at 20\,cm with projected baselines $\le 1170$\,m and data at 13\,cm
with projected baselines $\le 150$\,m.  The synthesized beam and rms
at 20\,cm were $19\farcs4 \times 4\farcs4$ and 0.19\,mJy\,beam$^{-1}$,
respectively.  At 13\,cm, the corresponding values were $15\farcs1 \times
3\farcs7$ and 0.17\,mJy\,beam$^{-1}$.  At each frequency, a map made from
the four consecutive phase bins clearly containing the pulsar yielded
a source consistent with the beam shape, with a pulse-averaged flux
density of 8.5\,mJy at both frequencies (hence implying a flat spectrum).
Images constructed from the data in the remaining 28 phase bins show no
source at the magnetar position.  This implies a $2\,\sigma$ limit to any
diffuse emission of $<4\%$ of the measured average pulsed flux.  Thus,
on scales of approximately $5''$ to $30''$ ($\sim\!0.1$ to $0.5$\,pc),
there is no evidence for any extended radio nebula surrounding \psr.
The $2\,\sigma$ limit on surface brightness implied by the 13\,cm data
is $\Sigma < 3.3\times 10^{-21}$\,W\,m$^{-2}$\,Hz$^{-1}$\,sr$^{-1}$.

\subsection{VLA Observations}

The fact that our first VLA followup observation at 3.6\,cm showed that
the source was 90\% linearly polarized was an early indication that
our initial hypothesis of an extended nebula was suspect.  No gated VLA
observations were made, and our best limit on diffuse emission associated
with the source is from our 2006 April 3 observations at 3.6\,cm first
reported in Camilo et al.\ (2006); roughly 75 minutes of on-source
data were recorded, yielding a map rms of $60\,\mu$Jy\,beam$^{-1}$.
The synthesized beam in the A-configuration for our $uv$ coverage had
a FWHM of $0\farcs34 \times 0\farcs17$.  Fitting a two-dimensional
Gaussian to the magnetar produced parameters wholly consistent with
this beam shape.  The source Gaussian plus a constant offset produced
flux densities of 0.26, 0.16, and 0.10\,mJy for nebula sizes of $1''$,
$2''$, and $4''$, respectively. Thus, on scales of $1''$ to $5''$ (0.015
to 0.1\,pc) excluding the pulsar location, any diffuse flux is $<3\%$
of the pulsed value of 9.1\,mJy on the date of the observation.

\subsection{VLBA Observations}

We can use the off-pulse data from the first-epoch VLBA observations to
set a stringent limit on the unpulsed radio flux from \psr\ on smaller
angular scales.  The rms of the off-pulse 3.6\,cm map at the magnetar
location is 55\,$\mu$Jy\,beam$^{-1}$ and the beam is 2.9\,mas$\times
1.7$\,mas which corresponds to a linear size of about 4\,AU. Since the
on-pulse source flux density was 8.0\,mJy on this day, this implies a
$3\,\sigma$ limit of $<2\%$ on any unpulsed emission associated with
the neutron star itself; this is a conservative upper limit, as any
miscentering of the pulse window as well as any emission far from the
main pulse (as is seen on many occasions; Camilo et al.\ 2007a) will act
to raise the rms in the off-pulse image. This result also rules out any
diffuse emission on scales of 5 to 50\,AU as bright as 0.05 to 5\,mJy,
although the latter is ruled out by the ATCA observations (\S~4.1)
which fail to detect a point source in the off-pulse data an order of
magnitude fainter.

Taken together, the three sets of observations both exclude significant
off-pulse emission from the magnetar's magnetosphere and limit the
fraction of flux in a diffuse surrounding nebula to $<4\%$ of the
pulsed flux on scales ranging from 5\,AU to 0.5\,pc.  The faintest
known pulsar wind nebulae (either isolated or inside a composite SNR)
have $\Sigma\sim\!3 \times 10^{-20}$\,W\,m$^{-2}$\,Hz$^{-1}$\,sr$^{-1}$
(see data in Green 2004).  The limits reported here on scales of
$\sim\!5''$--$30''$ (\S~4.1) are a factor of 10 below this level and
are therefore extremely constraining.

\section{The Velocity of XTE~J1810--197 and Implications for Magnetar Birth}

Prior to the discovery of radio emission from \psr, magnetar distance
estimates primarily were derived from X-ray absorption and the reddening
along the line of sight (Durant \& van Kerkwijk 2006), or from an
associated SNR or star cluster. The radio detection adds two distance
proxies that can be measured: a kinematical distance from H{\sc i}
absorption and the pulsar dispersion measure. Minter at al.\ (2007)
provide a current review of all these determinations, concluding that
the distance to \psr\ is $3.5 \pm 0.5$\,kpc. We adopt this value in
determining the transverse velocity of the magnetar.

Using the best value for the distance and the weighted average proper
motion value (Table~\ref{Table:fit}), the transverse velocity of the
magnetar is 223\,km\,s$^{-1}$ uncorrected for differential Galactic
rotation and solar motion.  We have computed these corrections for a
distance of $3.5 \pm 0.5$\,kpc; the velocity with respect to the local
standard of rest is $V_{\perp,{\rm LSR}} = 212 \pm 35$\,km\,s$^{-1}$
($1\,\sigma$).  In Galactic coordinates, the velocity vector has a
position angle of $89\fdg5$ --- i.e., its motion is directly along the
Galactic plane.

This transverse velocity is comparable to the mean for the population
of known normal radio pulsars ($246 \pm 22$\,km\,s$^{-1}$, according to
Hobbs et al.\ 2005) and far below that of the fastest ordinary pulsars
with direct proper motion and distance measurements ($V_{\perp} =
1100$\,km\,s$^{-1}$; Chatterjee et al.\ 2005).  For a random orientation
of the three-dimensional velocity vector, it is likely that the space
velocity is $\la\!400$\,km\,s$^{-1}$ ($2\,\sigma$).  Despite all the
striking characteristics of \psr, its velocity is completely ordinary.

We can conclude from the foregoing that the generation of magnetar
magnetic fields does not {\em require\/} that the forming neutron star
acquire an extremely high velocity.

Given the large excursions observed in the period derivatives of
magnetars, characteristic ages obtained from the spin parameters
are unreliable. Nonetheless, given the association of some AXPs
with SNRs, as well as the Galactic distribution of AXPs and SGRs,
it is generally thought that these objects are young ($<10^5$\,yr).
Given our well-constrained proper motion vector for \psr, we can
extrapolate back along its path for clues to its birth. There are no
objects of interest within a cone representing $2\times10^4$\,yr of motion
(its characteristic age in 2007 February), but the location at $\approx
10^5$\,yr ($\mbox{R.A.} = 18^{\mathrm h} 10^{\mathrm m} 38^{\mathrm s}$,
$\mbox{Decl.} = -19\arcdeg 24' 21''$) lies within the bounds of a very low
surface brightness shell of radio emission (barely visible in the MAGPIS
image\footnote{Available at http://third.ucllnl.org/cgi-bin/gpscutout .})
approximately $15'$ in diameter (15\,pc if at a distance of 3.5\,kpc). The
shell is not visible in either the MSX $20\,\mu$m image or the GLIMPSE
$8\,\mu$m image, suggesting it is non-thermal. Since such shells cover
a significant fraction of the Galactic plane near $b=0\arcdeg$, this
coincidence is at best weak evidence for an association with the magnetar.

The peak flux density of \psr\ decreased by about an order of magnitude
in late 2006 July and has shown no sign of recovery as of early 2007.
Unless the radio brightness returns to its levels of early 2006, it
will not be possible to measure a parallax.  If the velocity of \psr\ is
typical of the magnetar population, optical/IR and/or X-ray measurements
to determine the proper motions of other magnetars will be challenging,
but may still provide useful constraints.

\acknowledgements

The VLBA and VLA are telescopes operated by the National Radio Astronomy
Observatory, a facility of the National Science Foundation operated
under cooperative agreement by Associated Universities, Inc. The ATCA
is part of the Australia Telescope, which is funded by the Commonwealth
of Australia for operation as a National Facility managed by CSIRO.
This research was supported in part by grants from the NSF to DJH
(AST-05-07598) and FC (AST-05-07376); SC acknowledges support from the
University of Sydney Postdoctoral Fellowship Program.

\clearpage 

\end{document}